\documentclass[conference]{IEEEtran}

\usepackage{textcomp}
\usepackage{comment}
\usepackage{ifthen}
\usepackage{subcaption}
\usepackage{booktabs}
\usepackage{footnote}
\usepackage{graphicx}
\usepackage{paralist}
\usepackage{enumitem}
\usepackage{float}
\usepackage{color}
\usepackage{xspace}	
\usepackage{flushend}
\usepackage{balance}
\usepackage{enumitem}
\usepackage{marvosym}
\usepackage{tikz}

\def\BibTeX{{\rm B\kern-.05em{\sc i\kern-.025em b}\kern-.08em
    T\kern-.1667em\lower.7ex\hbox{E}\kern-.125emX}}

\newboolean{authnotes}

\ifthenelse{\boolean{authnotes}}
{
	\newcommand{\amy}[1]{\footnote{{\bf Amy: #1}}}
	\newcommand{\rajeev}[1]{\footnote{{\bf Rajeev: #1}}}
	\newcommand{\michele}[1]{\footnote{{\bf Michele: #1}}}
}
{
	\newcommand{\amy}[1]{}
	\newcommand{\rajeev}[1]{}	
	\newcommand{\michele}[1]{}
}

\newcommand{\tdma}{On-demand TDMA\xspace}
\newcommand{\lora}{LoRa\xspace}

\newcommand{\lorawan}{LoRaWAN\xspace}
\newcommand{\lbt}{Listen Before Talk\xspace}

\begin{document}

\title{On-Demand TDMA for Energy Efficient Data Collection with LoRa and Wake-up Receiver}

\author{\IEEEauthorblockN{
		\IEEEauthorrefmark{1}Rajeev~Piyare, 
		\IEEEauthorrefmark{1}Amy~L.~Murphy, 
		\IEEEauthorrefmark{2}Michele~Magno, and  
		\IEEEauthorrefmark{2}Luca~Benini }
	
	\IEEEauthorblockA{
		\IEEEauthorrefmark{1}Fondazione Bruno Kessler, Trento, Italy \{piyare, murphy\}@fbk.eu} 
	\IEEEauthorrefmark{2}Integrated Systems Laboratory, ETH Z\"{u}rich \{michele.magno, lbenini\}@iis.ee.ethz.ch
}

\maketitle

\begin{abstract}
	
Low-power and long-range communication technologies such as LoRa are becoming popular in IoT applications due to their ability to cover kilometers range with milliwatt of power consumption. One of the  major drawbacks of LoRa is the data latency and the traffic congestion when the number of devices in the network increases. Especially, the latency arises due to the extreme duty cycling of LoRa end-nodes for reducing  the overall energy consumption. 
To overcome this drawback, we propose a heterogeneous network architecture and an energy-efficient On-demand TDMA communication scheme improving both the device lifetime and the data latency of standard LoRa networks. We combine the capabilities of micro-watt wake-up receivers to achieve ultra-low power states and pure asynchronous communication together with the long-range connectivity of LoRa. Experimental results show a data reliability of 100\% and a round-trip latency on the order of milliseconds with end devices dissipating less than 46~$mJ$ when active and 1.83~$\mu W$ during periods of inactivity, lasting up to 3 years on a 1200~$mAh$ Lithium battery.
 
\end{abstract}



\section{Introduction}
\label{sec:introduction}

The recent Internet of Things wave is boosting the development of connected smart wireless low-power devices. The most common IoT end devices are sensor nodes that measure physical properties such as vibration, pressure, and temperature. These sensor nodes are usually battery powered and thus  have limited lifetime requiring careful power management as replacing or recharging batteries frequently will not only be costly but infeasible in many application scenarios.  Thus, the design of wireless sensing devices and applications pose several challenges, especially in communication and networking. In fact, the wireless transceiver is usually one of the most power hungry subsystems consuming around tens of milliwatts. The problem is exacerbated when there are many nodes in the network and the required range is on the order of hundreds of meters or even kilometers. 
 
Applications such as smart city and environment monitoring have pushed the research towards novel long-range wireless technologies like LoRa, SigFox, and NB-IoT. Among these,  LoRa\texttrademark~ is constatly increasing its success in the market due its open protocol standard and Chirp Spread Spectrum (CSS) modulation technique that allows recovering data from weak signals even below the noise floor. Several solutions for the LoRa MAC layer have been proposed such as Listen Before Talk (LBT)~\cite{Symphony} while the \lorawan is the most popular one. \lorawan is based on the ALOHA protocol where the distributed battery-operated end devices communicate directly to an always-on gateway when they have data ready to send, making it ideal for applications that transmit sporadically in small bursts with a low-traffic. The energy efficiency comes by duty-cycling the radios when they are not transmitting.    

\lorawan is mainly designed for uplinks from the end devices to the gateway. Downlink communications from the gateway to the node at any other time must wait until the next uplink transmission. This discourages any fine-grained scheduling and or coordination of transmissions by the gateways. 
Several studies have evaluated the scalability of ALOHA using simulations and concluded that although ALOHA operates well under low-traffic loads, it suffers from uplink traffic congestion due to its inability to check whether the channel is busy before transmitting~\cite{Bor:mswim, pop2017does, goodbye-aloha}. Moreover,  duty-cycling restrictions in \lorawan impose a trade-off between downlink traffic, latency, and power consumption~\cite{Watteyne2017}. This may affect the performance of applications like structural health and seismic activity monitoring where both low-latency and low-power are required.  

Wake-up receiver has the capability to continuously monitor the wireless channel while consuming power orders of magnitude lower than commodity radio hardware typically utilized in wireless sensor platforms~\cite{piyare2017ieee}. To benefit from this technology, sensor nodes are equipped with wake-up receiver circuitry and put into low-power modes, waiting for a remote trigger signal. Upon detection of the valid wake-up beacon, it triggers the main node out of sleep mode to exchange data ``instantly", thus reducing latency. Recent WuR designs also perform address matching with micro-watts of power avoiding false wake-ups~\cite{magnoITII}. These features of wake-up receiver allows ``\textit{pure}" asynchronous communication by activating the system on-demand. Since wake-up receiver detects signals with very low-current, it significantly reduces the wasteful power of idle listening, improving energy efficiency without compromising data latency. 

In this work, we propose a  novel receiver-initiated \emph{\tdma}  MAC protocol that provides a deterministic performance improving reliability and network efficiency of LoRa. To achieve this, we exploit state-of-the-art ultra low-power wake-up receiver in-combination with LoRa. 
In summary, this work provides following contributions:
 \begin{enumerate}[label=(\roman*)]
 	\item a network architecture that leverages wake-up receiver and LoRa for enabling low-latency and energy efficient data collection for LoRa networks (\S\ref{sec:network}).
 	\item design and implementation of an \tdma MAC that exploits dual-radios for managing channel access and packet collisions (\S\ref{sec:mac}).
 	\item performance evaluation of the complete system using an indoor office testbed. Our results show that \tdma achieves an improvement of at least 1.72$\times$ in terms of latency and extended node lifetime of 1.4$\times$ with 100\% system reliability over Listen Before Talk MAC scheme for LoRa (\S\ref{sec:results}). 
 \end{enumerate}


\section{Energy efficient data communication network architecture and protocol}

In the following subsections, we outline our proposed network architecture and  \tdma protocol for LoRa communication to achieve the goal of improving energy efficiency without compromising data latency and network reliability. The proposed network architecture is designed to realize many-to-one and one-to-many communication.

\subsection{Heterogeneous network architecture}
\label{sec:network}
We propose a heterogeneous IoT network where the end devices are equipped with a dual-radio interface: a wake-up receiver (WuRX) for short-range networking and a LoRa transceiver for long-range communication. \emph{We target applications where the the sink needs to pull data from the nodes upon demand. For instance, the gateway querying a set of cluster nodes on a wine farm for the latest measurements.} These applications are quite popular in IoT~\cite{BRUNLAGUNA}.

\begin{figure}[t!]
	\centering
	\includegraphics[width=0.8\linewidth]{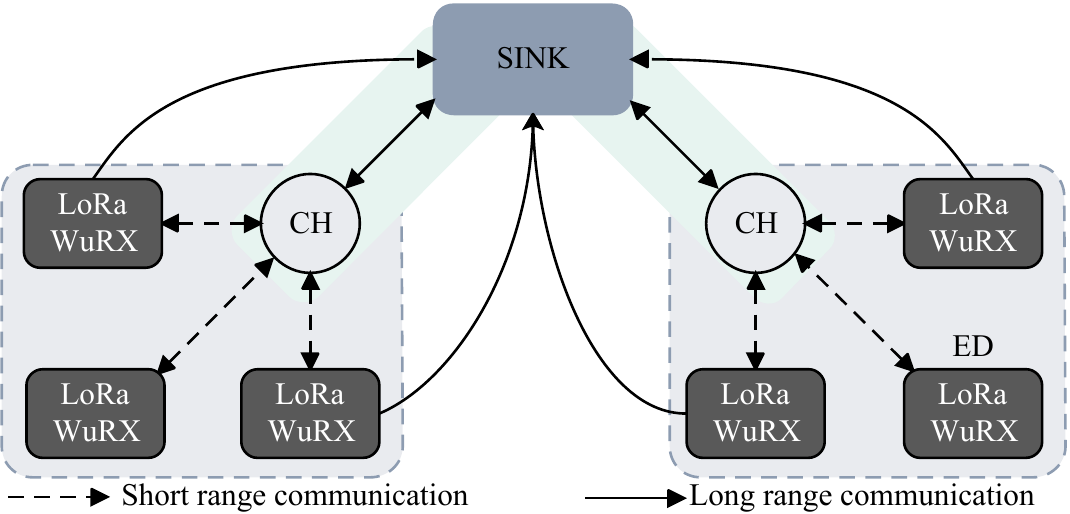}
	\caption{Heterogeneous IoT topology for sensor networks.}
	\label{fig:topology}
	\vspace*{-4.0ex}
\end{figure}

With pure asynchronous communication over the WuRX, the end devices are not periodically or continuously listen to the channel, overcoming the issues of idle listening and latency, further improving the energy efficiency. Unlike \lorawan, where a gateway communicates directly to the nodes, we partition the network into clusters. The clusters are comprised of sensor and actuator nodes forming a star topology, similar to that of \lorawan, but any down-link communication from the gateway to the nodes (one-to-many) must go through the cluster heads due to the short communication range of wake-up receivers.  In detail, the whole network is composed of three different nodes as illustrated in Fig.~\ref{fig:topology}:

\begin{enumerate} 
	\item \emph{End device (ED)} responsible for sensing and is equipped with both, a wake-up receiver and a \lora transceiver and reside in a low-power mode.
	\item \emph{Cluster head (CH)} is in-charge of relaying commands from the sink to the EDs. CHs are also equipped with both radios. Each CH is assigned an unique ID allowing the sink to query each CH at a time, thus reducing the interference from other clusters.
	\item \emph{Sink} acts as a gateway and has no energy constraints. Therefore, the sink can be always-on and listening for any incoming data. Unlike ED and CH, the sink only offers long-range communication, i.e, only LoRa radio.
\end{enumerate}

The transmission between the sink and the CH is bidirectional (Sink$\longleftrightarrow$CH), allowing the gateway to update network parameters or collect data from the CHs directly. CH and the EDs also communicate bidirectionally (ED$\longleftrightarrow$CH) so that the CH can trigger the EDs and the EDs can query neighboring nodes without passing through the gateway every time. On the other hand, EDs communicate unidirectionally to the sink (ED$\longrightarrow$Sink). This reduces delays as the nodes need not hop the data over the CH to get back to the sink i.e., single-hop communication as done in LoRaWAN (many-to-one).

\subsection{On-demand TDMA  MAC Layer}
\label{sec:mac}
One of the issues faced by \lorawan is packet collisions as the  nodes may try to transmit data concurrently upon wake-up causing network congestion. To mitigate this, we have implemented a low-complexity yet robust slot scheduler atop receiver-initiated MAC, thus the name \textit{ \tdma}. The slot assignment per node allows full utilization of the bandwidth while reducing the channel activity and eliminating packet collisions in the network.  In all cases, our goal is to pull data from all the nodes in the most energy efficient and low-latency manner. Therefore, the core of the proposed MAC is receiver-initiated asynchronous communication over wake-up receivers providing network-wide data collection with reduced packet collisions. Next, we present this \tdma scheme and its inner workings.

\begin{figure}[t!]
	\centering
	\includegraphics[width=0.9\linewidth]{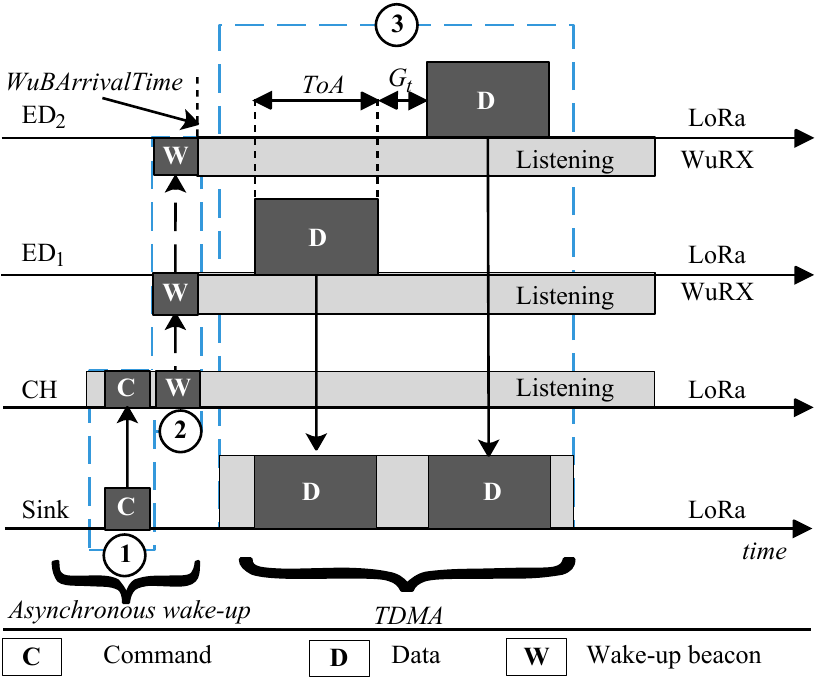}
	\caption{Operation of Receiver-initiated On-demand TDMA.}
	\label{fig:tdma}
	\vspace*{-4.0ex}
\end{figure}

The proposed MAC requires three steps to operate efficiently. Fig.~\ref{fig:tdma} exemplifies the operation of On-demand TDMA using short-range and long-range radios.  The first step is performed by the sink which can be located a few kilometers away from the clusters. \textcircled{\small{1}}~Sink node sends the command to the specific cluster head to initiate data collection. \textcircled{\small{2}}~The CH then broadcasts a wake-up beacon for activating all the EDs within the cluster. The use of broadcast beacon is two-fold. First, broadcasting allows triggering all the nodes in range reducing latency w.r.t to the unicast where several addressed wake-up beacons are transmitted for activating each specific node. Second, transmitting a single beacon saves energy both at the sink and the CH side as multiple beacon transmissions are expensive. Next, transmitted wake-up beacon serves two purposes: 
\begin{inparaenum}[\em i)]
	\item triggers the end devices from sleep to active state using the wake-up receiver.
	\item For proper operation of the \tdma all nodes in the cluster need to agree on the  time slots, therefore, clock synchronization is required. The wake-up beacon provides a fine-grained time synchronization among the EDs achieved by the asynchronous network wake-up without requiring considerable amount of down-link resources. The average time synchronization among the nodes is measured to be 95~$\mu$s. 
\end{inparaenum}
\textcircled{\small{3}}~The final step is performed by the end devices.  All the EDs are now active and synchronized.
First, each ED enters into a slot reservation phase where the EDs occupy the slot according to their ID, $N_{id}$ i.e., the node with the ID:1 occupies the first slot while the node with the maximum ID occupies the last slot. The ED measures the exact time when it was triggered by the wake-up beacon sent by the CH, $WuBArrivalTime$. Each node then computes the start of their time slot from the  $WuBArrivalTime$ as:
\begin{equation}
\label{eq:slot}
T_{NextSlot} = WuBArrivalTime + (ToA_{pkt} + G_{t}) N_{id}
\end{equation}

The slot size is determined by computing the time on air, $ToA_{pkt}$  for the LoRa data packet depending on the payload size with a pre-defined guard time, $G_{t}$ of 6~ms. The guard time guarantees that the  window is large enough for the transmission and compensates for clock drift, which may be detrimental with an increasing number of EDs.  
Finally, the EDs start transmitting the data packets directly to the sink over the LoRa module as per the slot schedule illustrated in Eq.~(\ref{eq:slot}). In our initial implementation, slot allocation is statically defined a priori at network configuration time.
Through this on-demand  energy effficient scheduling approach, the packet collisions in the network is greatly reduced. Not only this, but also the lifetime is improved  up to 1.4$\times$ as nodes spend most of the time in deep sleep until queried by the gateway. We next evaluate the proposed MAC  through an indoor testbed experiments.

\section{Performance Evaluation}
\label{sec:results}
This section presents the in-field results of our evaluation of the On-demand TDMA protocol through a test-bed analysis. 

\subsection{Scenario and settings}
To show the proof of concept, we deployed a total of 11 wireless sensor nodes in an indoor office environment based on the dual-radio platform previously developed by some of the authors of this work~\cite{magno2017wulora}. Of the 11 nodes, 9 are designated as end devices responsible for sensing, one acts as the cluster head, and one as a sink. We have used the SX1276 radio module from Semtech for data and wake-up beacon transmission.
To use the SX1276 as a wake-up transmitter, it is configured for transmission using OOK modulation. The wake-up beacon is 2B long and are transmitted at 1~kbps. The SX1276 consumes 50mW in listening mode and 250mW when transmitting at output power of +14dBm. The wake-up receiver consumption is  1.83~$\mu$W in listening mode and 284~$\mu$W when it is decoding the address. All experiments were performed with an 8B application payload with the $G_{t}$ of 6~ms.  

\begin{figure*}[t]
	\begin{center}
		\begin{subfigure}{0.3\linewidth}
			\includegraphics[width=\linewidth]{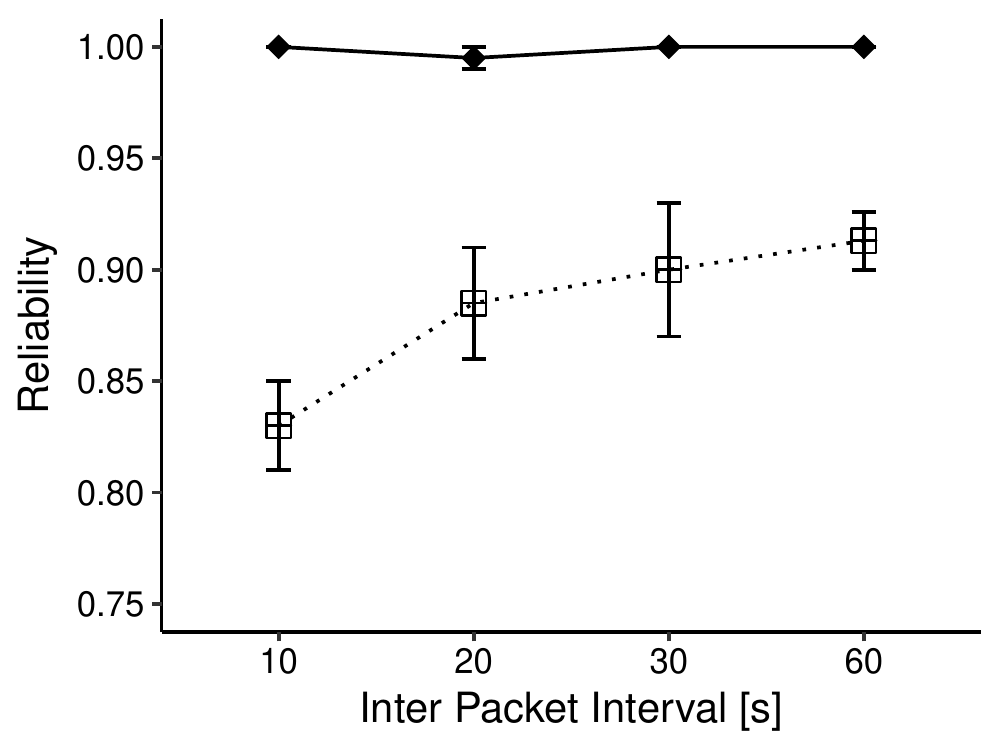}
			\caption{Reliability}
			\label{fig:pdr}
		\end{subfigure}
		\begin{subfigure}{0.3\linewidth}
			\includegraphics[width=\linewidth]{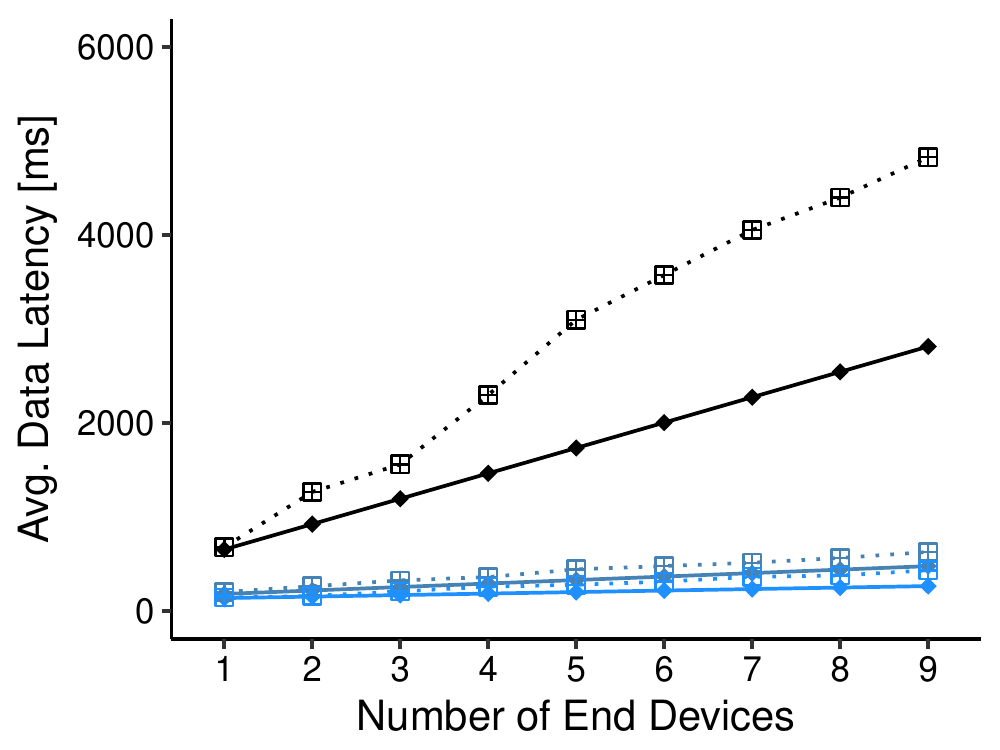}
			\caption{End-to-end latency}
			\label{fig:latency}
		\end{subfigure}
		\begin{subfigure}{0.38\linewidth}
			\includegraphics[width=\linewidth]{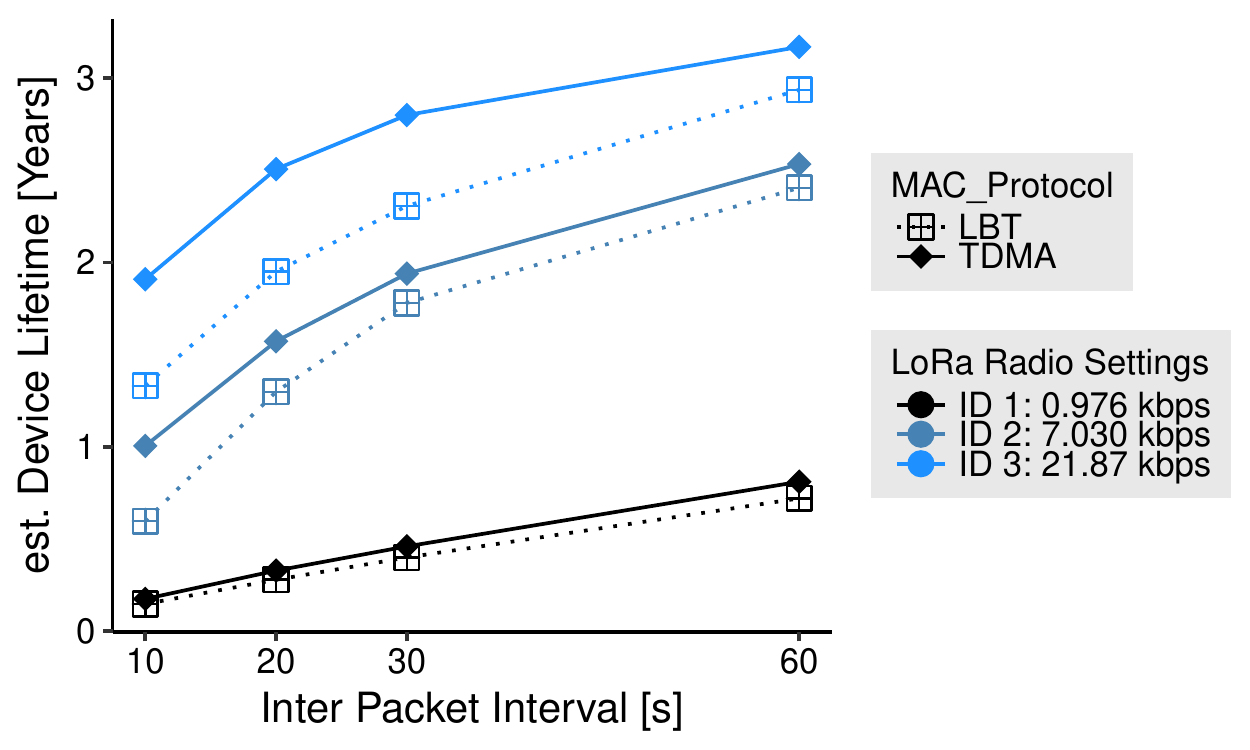}
			\caption{Device lifetime}
			\label{fig:lifetime}
		\end{subfigure}
		\caption{Network performance of \tdma w.r.t to \lbt (LBT) MAC.  }
		\label{fig:performance}
	\end{center}
\vspace*{-4.0ex}
\end{figure*}

In \lora networks, switching to different data rate affects the receiver sensitivity i.e, at low data rate, \lora packets can be received at much longer range with high reliability. To explore this, we selected three different \lora radio settings. With setting \emph{ID~1}, we chose the most robust transceiver setting with the lowest data rate of 0.976~kbps and spreading factor of 12 leading to the time on air of 264~ms. Setting\emph{ ID~2} represents a midrange data rate of 7.03~kbps with time-on-air of 31~ms while setting \emph{ID~3} represents the shortest airtime of 9~ms with bit rate of 21.87~kbps.

In contrast to LoRaWAN, all our experiments use single channel for LoRa communication. Therefore, to provide a fair comparison we have  evaluated performance of the \tdma against Listen Before Talk (LBT) protocol, a single channel MAC with random backoff. The random backoff interval for LBT is restricted to be between \{0, 2s\}. 

\subsection{Network Performance Analysis}
The effectiveness of our solution is evaluated by investigating the following performance metrics. All metrics have been computed based on 300 data packets exchanged between the nodes. For all the experiments, we vary the network load by exploring a range of inter-packet intervals (IPI) from 10s to 60s.

\subsubsection{Packet delivery ratio}
We begin our evaluation by considering the data collection reliability of the network in terms of generated data packets and those successfully received by the sink. We first observe the PDR of 100\% for \tdma for varying IPIs (Fig.~\ref{fig:pdr}). The higher PDR is attributed to the facts that:
\begin{inparaenum}[\em i)]
	\item  each ED transmits in its own allocated slot, reflecting the ability to handle higher bandwidth without collisions. 
	\item the maximum distance between the nodes in our experiments was 10m, therefore, all the trigger samples were correctly received by the wake-up receivers.
\end{inparaenum}

On the other hand, the average PDR for the LBT protocol varies between 83\% and 91\% for varying network traffic.
This is because the channel activity detection (CAD) feature similar to CSMA is only useful to detect LoRa preamble symbols. As CAD is not able to detect all on-going transmissions, especially when the preamble has been already sent, this causes packet collisions at the receiver, dropping some packets. 

\subsubsection{End-to-end data latency}
Next, we turn our attention to the network latency as captured in Fig.~\ref{fig:latency}. We compare the three different \lora radio settings and its impact on the overall network latency using both the protocols. For latency, the \textit{round-trip time} is computed as the difference between the sink node transmitting the command packet and receiving all the data from the nine end devices. 
For both protocols, the latency increases linearly with the increasing number of EDs for each radio setting.
This delay includes the wake-up delay measured at 17~ms, which is consistent with the parametrization of the wake-up receiver. For higher number of nodes, however, \tdma compares favorably to LBT because nodes are able to send packets without collisions in their own slots. The only extra delay that occurs per transmission is the 6~ms guard time used to compensate for clock drifts.
On the other hand, nodes using LBT must compete for the medium and backoff whenever the other node is transmitting. The competition for medium is low with fewer end devices but as the network size increases, so does the latency due to frequent backoffs.  For the highest data rate scenario with the shortest packet transmission time i.e, Setting ID~3, the latency of LBT is $1.65\times$ higher than \tdma and the performance difference is even more significant for settings 1 and 2 (slower data rate), with more nodes and longer transmission time. For instance, to collect data from nine nodes, LBT requires $1.72\times$ longer than \tdma for radio setting ID~1.

\subsubsection{Energy efficiency}
We also estimate the end device lifetime to compare the performance of the \tdma against LBT. Device lifetime is a critical metric as it directly affects the network lifetime. The evaluated sensor node draws an average current of 0.56~$\mu$A leading to a  theoretical standby time of 244 years on a 1200~mAh Lithium Polymer battery, only if batteries could last that long. Next, we estimate the end device lifetime when it is actively participating in data collection rounds with varying packet intervals. As seen in Fig.~\ref{fig:lifetime}, nodes employing \tdma scheme, can last up to 3 years when polled every minute for data collection. This directly translates into energy saving due to a low radio-on time required by \tdma compared to LBT to transmit the same amount of payload. Even with shorter trigger intervals, i.e., every 10~s, \tdma provides lifetime improvement of up to $1.4\times$ in contrast to LBT. On the other hand, the lifetime of the end node is less than a year for the lowest data rate setting  w.r.t both protocols because of the extra power consumed when the radios are active for long period. Overall, the \tdma demonstrates significant gains in energy that can be achieved over channel sensing schemes for long-range networks. \smallskip


\section{Conclusions}
\label{sec:conclusions}

In this paper, we proposed an asynchronous TDMA based MAC  protocol for short- and long-range networks. \tdma provides efficient broadcast service for data collection and synchronization improving the performance of LoRa networks. The time  synchronization on the order of tens of microseconds is achievable using wake-up radios without requiring complex synchronization algorithms.
It has been shown from testbed experiments that \tdma significantly improves system scalability and energy efficiency by offering network reliability of 100\%  with end devices dissipating 1.83$\mu$W of power during periods of inactivity. We also observed that different LoRa transceiver settings can have significant variations in airtime for a LoRa data packet. Thus, the selection of communication parameters has a tremendous impact on the scalability of a LoRa deployment. Our proposed protocol supports a node wake-up delay on the order of milliseconds with a round-trip latency below a second through a 2-hop network while sustaining nodes for up to 3 years.


\balance
\bibliographystyle{IEEEtran}
\bibliography{bib}

\end{document}